\documentclass[conference]{IEEEtran}
\IEEEoverridecommandlockouts
\usepackage{cite}
\usepackage{amsmath,amssymb,amsfonts}
\usepackage{algorithm}
\usepackage{textcomp}

\usepackage{mathtools}
\usepackage{algpseudocode}
\usepackage{graphicx}
\usepackage{url}
\usepackage{textcomp}
\usepackage{xcolor}
\definecolor{airforceblue}{rgb}{0.36, 0.54, 0.66}
\definecolor{applegreen}{rgb}{0.55, 0.71, 0.0}
\definecolor{ao(english)}{rgb}{0.0, 0.5, 0.0}
\definecolor{ao}{rgb}{0.0, 0.0, 1.0}
\usepackage{caption}
\usepackage{subcaption}
\usepackage{enumitem}

\def\BibTeX{{\rm B\kern-.05em{\sc i\kern-.025em b}\kern-.08em
    T\kern-.1667em\lower.7ex\hbox{E}\kern-.125emX}}

\begin{document}
\title{Future-Proofing Mobile Networks: A Digital Twin Approach to Multi-Signal Management}

\author{\IEEEauthorblockN{
Roberto Morabito\IEEEauthorrefmark{1},
Bivek Pandey\IEEEauthorrefmark{3}, 
Paulius Daubaris\IEEEauthorrefmark{3},\\ 
Yasith R Wanigarathna\IEEEauthorrefmark{3},
Sasu Tarkoma\IEEEauthorrefmark{3}}
\IEEEauthorblockA{\IEEEauthorrefmark{1}Department of Communication Systems, EURECOM, France.}
\IEEEauthorblockA{\IEEEauthorrefmark{3}Department of Computer Science,
University of Helsinki, Finland.}

\thanks{
A shortened version of this paper is currently under review for publication in an IEEE magazine. If accepted, the copyright will be transferred to IEEE.
}
}

\maketitle
\pagestyle{plain}

\begin{abstract} 
Digital Twins (DTs) are set to become a key enabling technology in future wireless networks, with their use in network management increasing significantly. We developed a DT framework that leverages the heterogeneity of network access technologies as a resource for enhanced network performance and management, enabling smart data handling in the physical network. Tested in a \textit{Campus Area Network} environment, our framework integrates diverse data sources to provide real-time, holistic insights into network performance and environmental sensing. We also envision that traditional analytics will evolve to rely on emerging AI models, such as Generative AI (GenAI), while leveraging current analytics capabilities. This capacity can simplify analytics processes through advanced ML models, enabling descriptive, diagnostic, predictive, and prescriptive analytics in a unified fashion. Finally, we present specific research opportunities concerning interoperability aspects and envision aligning advancements in DT technology with evolved AI integration.

\end{abstract}

\maketitle

\section{Introduction}

In recent years, DTs have gained significant attention for their ability to create digital replicas of entire systems, enhancing system monitoring, optimization, and predictive maintenance\cite{Singh2021}. Such systems have been used in a multitude of domains, for example, manufacturing, robotics, healthcare, and urban planning, demonstrating their versatile applications and impact across various industries\cite{Mihai2022}.
While the application of DTs in these areas has been highly successful, the field of mobile networks is beginning to see significant advancements and growing importance in the implementation and use of DT technology, especially in the path towards 6G \cite{lin20236g}.

This potential is exemplified with the interest within the 3rd Generation Partnership Project (3GPP) forum, where discussions are focused on leveraging DTs to advance network management and orchestration, highlights their critical role in upcoming releases. These discussions identify numerous challenges and opportunities that necessitate further Research and Development (R\&D) efforts \cite{3gppTS28915}.
Such focus on DTs is expected to culminate in the concept of 'Massive Twinning', indicating that 6G will extensively integrate DT technology\cite{hexa2021deliverable}, to provide data-driven insights aimed at improving communication efficiency and network performance.

As we move towards the realization of future mobile networks, enhancing communication efficiency is particularly important for all those smart environment use cases where the co-existence of heterogeneous network access technologies (i.e., 3GPP \textit{and} non-3GPP technologies) represents a default operating condition. In such scenarios, end-devices can leverage multiple radio access technologies (RATs) and opportunistically choose the most suitable one based on the type of application being used or the current network status, guided by AI-powered analytics to optimize communication methods.

Within this context, we envision the necessity of enhancing the current state of the art of DT solutions. While the existence of DT solutions for wireless networks has been increasing lately \cite{Ericsson2022, Vodafone2022}, there remains a gap that presents a significant opportunity. Specifically, the heterogeneity of devices and network access technologies is a resource that can be leveraged for future mobile networks, yet existing DT solutions do not fully capitalize on this potential.
Additionally, we identify another opportunity in the evolution of current data-driven approaches for DT \cite{Hui2023}, as we expect these will likely evolve to rely more and more on novel GenAI models \cite{chai2024generative}. This evolution can be particularly advantageous in ``smart'' scenarios, where the co-existence and integration of different data sources can generate more insightful analytics, thereby enhancing the overall capability of mobile networks while also considering the surrounding environment.

Given these premises, we present a DT framework that utilizes information from various network access technologies for advanced analytics and network management. We contextualize the system within a dynamic \textit{Campus Area Network}, comprising heterogeneous devices and network access technologies, where the DT provides a holistic overview of the network. Additionally, we discuss the use of GenAI and its potential to streamline and advance the analytics component of the DT. Finally, we delve into opportunities related to DT, such as interoperability and the integration of GenAI.  

\section{The DT Framework Design}
Building on the context of heterogeneous mobile networks discussed earlier and the potential key role of DT systems, we introduce our DT framework designed to accommodate the requirements of such environments. This framework aims to lay the foundation for DTs in multi-connectivity scenarios, where various RATs co-exist. 

\begin{figure}[!t]
    \centering
    \includegraphics[width=0.47\textwidth]{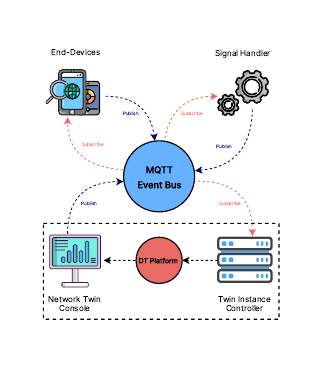}
    \caption{High-level architecture of the proposed system, illustrating the DT components within the squared dotted box, the physical network (End Devices), the Signal Handler, and the event bus for data transfer.}
    \label{fig:high-level-architecture}
\end{figure}

Figure \ref{fig:high-level-architecture} provides a high-level overview of the proposed DT framework. 
To achieve our goal of developing a system that supports heterogeneous mobile networks, we identified several key components: the physical network End-Devices (e.g., IoT devices, mobile devices), the Signal Handler, the Twin Instance Controller, and the Network Twin Console.
In a nutshell, the physical network consists of 
devices that act as primary data sources by publishing their signal information. The Signal Handler subscribes to this data, processes it, and publishes the refined information. The Twin Instance Controller retrieves the processed data and maintains a connection with the DT platform, ensuring real-time updates and synchronization. Finally, the Network Twin Console offers a way to visualize and interact with the gathered data, providing opportunities to perform analytics and optimizations on the physical twin.

A crucial requirement for this type of DT framework is its modular and adaptable architecture, which facilitates the integration with various DT platforms. While our current system interfaces with a well-defined vendor platform, the conceptual goal is to design dedicated interfaces, such as APIs, that ensure interoperability with other platforms (more about this in the last section), including those provided by major vendors, as well as open-source solutions.
 
\begin{figure*}[!t]
    \centering
    \includegraphics[width=0.95\textwidth]{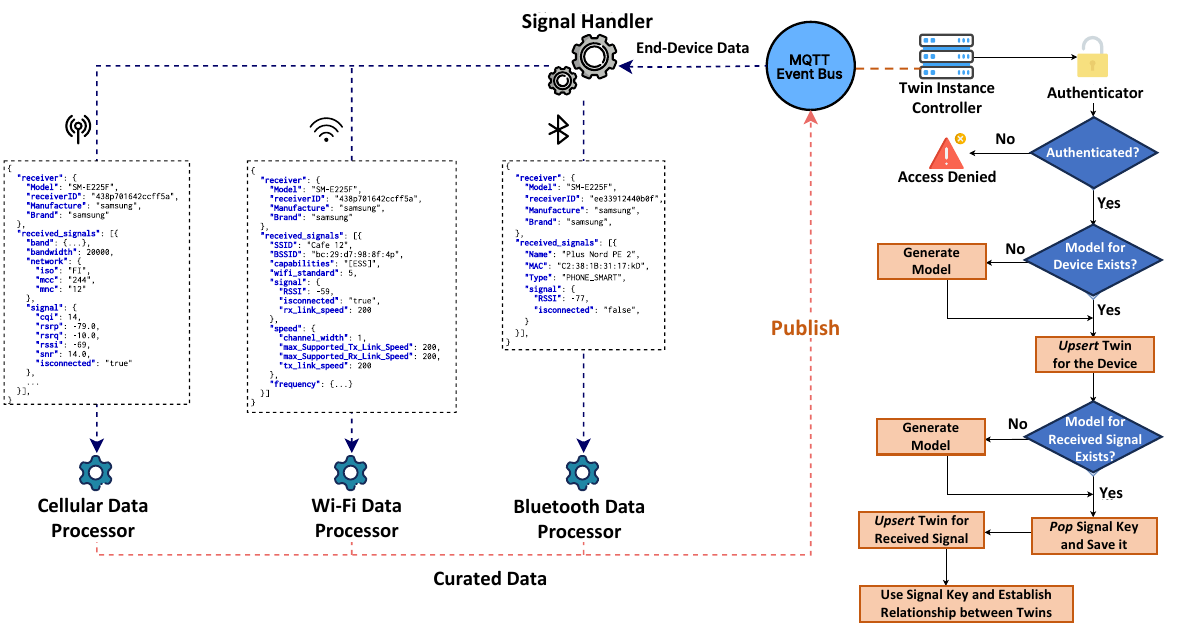}
    \caption{Workflow of the Signal Handler and Twin Instance Controller components. Incoming signals from different wireless technologies are processed before being published to the event bus. Once published, a digital representation is created.}
    \label{fig:signals}
\end{figure*}

In the following, we delve deeper into each of these components and their roles within the framework. 
\\
\\
\textbf{End-Devices as Data Source.} Each end-device embeds an application that records different types of incoming signals. The retrieval of such information is periodic and can be fine-tuned based on the specific needs. The high-level data structure for all signals follows a JSON schema, although the details describing the signals vary due to the unique characteristics of the originating devices and the data required. The JSON schema includes key fields such as the device model and various signal-specific metrics. Figure \ref{fig:signals}  illustrates the different types of information recorded for cellular, Wi-Fi, and Bluetooth signals. The retrieved data is published via an MQTT event bus for further processing by the Signal Handler. We chose an event bus approach to facilitate efficient handling of data streams. For our use case scenario, this approach is sufficient. However, in hyper-scale networks, where data collection can be extensive, a fully asynchronous and reactive model would be ideal to minimize the need for pooling.
\\
\\
\textbf{Signal Handler.} The Signal Handler is a key component of the framework, responsible for gathering, processing, and cleaning the data to be then published. The signal handling process is illustrated in the left side of Figure \ref{fig:signals}. This component handles incoming signals by spawning dedicated \textit{Data Processors} tailored to the type of signal received. This approach ensures easier management, as each processor can operate independently and start processing as soon as signal data is published and retrieved through the event bus. In fact, given the variety and volume of signal data incoming from multiple devices, using specialized processors for each signal type helps maintain organized and effective data processing. Curated data is then published to be utilized by other components of the DT framework. We have implemented this component to be flexibly executed either on the end-device itself or in a similar computing environment where the rest of the DT framework is executed. Based on the empirical analysis done in our related work \cite{pandey2024development}, hosting the Signal Handler in the cloud proved to be more effective, but this result might vary based on the type of end-device we consider (in our case, a mid-range smartphone). In cases of more resourceful devices or based on the type of processing required, we might be able to fully execute this locally on the end-device. In any case, our framework is equipped with the capability to opportunistically place the execution of this component based on the computing and data requirements.
\\
\\
\textbf{Twin Instance Controller.} The Twin Instance Controller is responsible for constructing and maintaining the digital representation of the network, including the devices, incoming signals and their characteristics. The right part of figure \ref{fig:signals} illustrates the workflow of such component. It begins by performing authentication to the DT platform. If authentication is successful, it checks whether the end-device model exists. If the model does not exist, the Twin Instance Controller requests the DT platform to create and instantiate a twin model. Otherwise, it updates the existing twin instance. A similar workflow is employed for the received signals. After generating the twin instances, relationships are established to show the connections between them. These relationships also store signal strength information to highlight the variability in signal reception across different devices within the DT system.
The Twin Instance Controller is responsible for maintaining the latest state information of the twin instances and relationships within the DT system. Each time it receives new data, it compares it with the current information available on the DT system to make the necessary adjustments. These adjustments can involve adding, removing, or updating twin instances or relationships. By ensuring real-time updates and synchronization, the Twin Instance Controller component plays a key role in accurately reflecting the state of the physical network within the DT environment.
\\
\\
\textbf{Network Twin Console.} 
The final component is a custom-built tool designed to represent devices and their captured signals in a user-friendly manner. This console is specifically tailored to plug into existing DT platforms, providing a comprehensive view of different models, twin instances, and their relationships. In addition to display the devices and their captured signals, the interface also shows the signal sources, as well as visualizing different key performance indicators (KPIs), such as time series for RSSI and RSRP, and provides an in-depth overview of the physical network under analysis. Furthermore, it can offer the opportunity to perform analytics and optimizations on the physical twin. These analytics support applications that implement the feedback loop of the digital twin system, transforming actionable insights into actions for the physical twin.

\section{Campus Area Network Use Case}
\begin{figure*}[!t]
    \centering
    \includegraphics[width=0.9\textwidth]{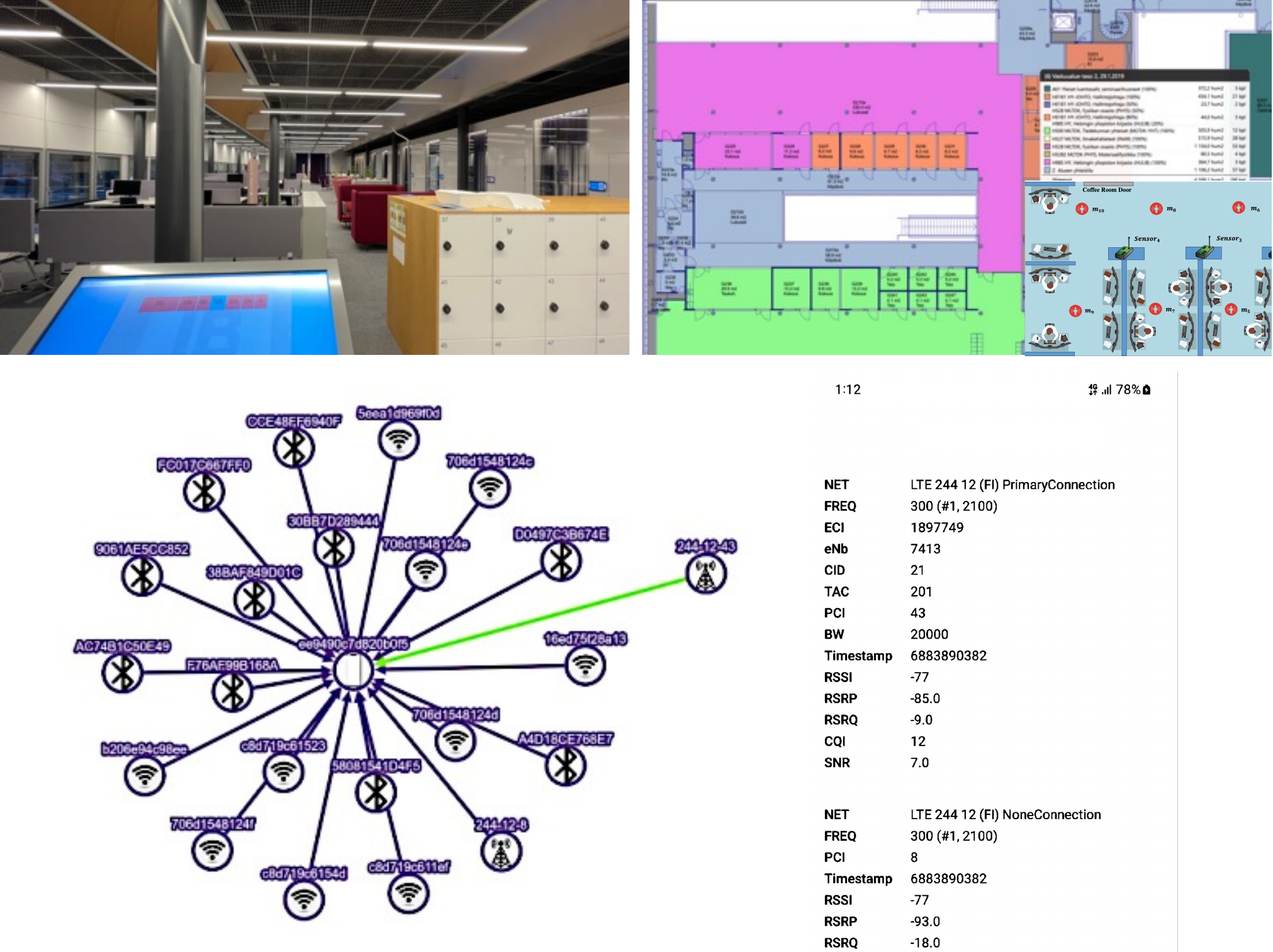}
    \caption{On top, overview of UbiKampus environment. The top-left image depicts the physical workspace with smart devices and interactive displays, and the top-right image shows the floor plan with monitoring points. On the bottom, the left image shows the network configuration sensed by a single smartphone device, while the right image presents the user interface of the application running at the end-device.}
    \label{fig:dtgraph}
\end{figure*}

Our proposed DT framework is being tested in an open working space at the University of Helsinki. 
UbiKampus spans a thousand square meters of open-plan office space and it serves as a living research laboratory, providing a dynamic environment where dozens of researchers work daily, and many visitors use the meeting rooms and shared areas. 
The space is equipped with an array of sensors and devices, including motion detectors, air quality sensors, and various other sensing devices that monitor temperature and carbon dioxide levels \cite{RintaHomi2021}. The upper part of Figure \ref{fig:dtgraph} provides a representation of this environment. The top-left image shows the physical workspace, equipped with smart devices and interactive displays that provide real-time data to users. The top-right image displays a floor plan of the UbiKampus area with indicators of various monitoring points. 

We consider UbiKampus to be equivalent to the concept of a \textit{Campus Area Network} and envision it as the university's equivalent of a personal area network but scaled for a campus environment. 


The open network at UbiKampus allows for seamless connectivity of smart devices, supporting our goal of testing the DT framework due to the availability of diverse RATs (e.g., 5G, Bluetooth, WiFi) and heterogeneous connected devices. 

The reason for testing our framework in such an environment is that having a full overview of a network and all its different capabilities can enable us to develop advanced analytics. These analytics can then develop algorithms to select the most appropriate communication interface for sending sporadic messages, taking into account both the device status and the characteristics of the data to be sent. 
Moreover, the framework enables the implementation of resource management and network optimization policies tailored for multi-connectivity networks. This is crucial in an environment like UbiKampus, where numerous devices and sensors operate simultaneously, requiring smart management to maintain optimal performance. 
The variety of sensors and devices at UbiKampus provides a rich source of data, allowing us to test and enhance these mechanisms in a real-world setting.


\section{The DT Framework in Action}
 \begin{figure*}[!t]
    \centering
    \includegraphics[width=\textwidth]{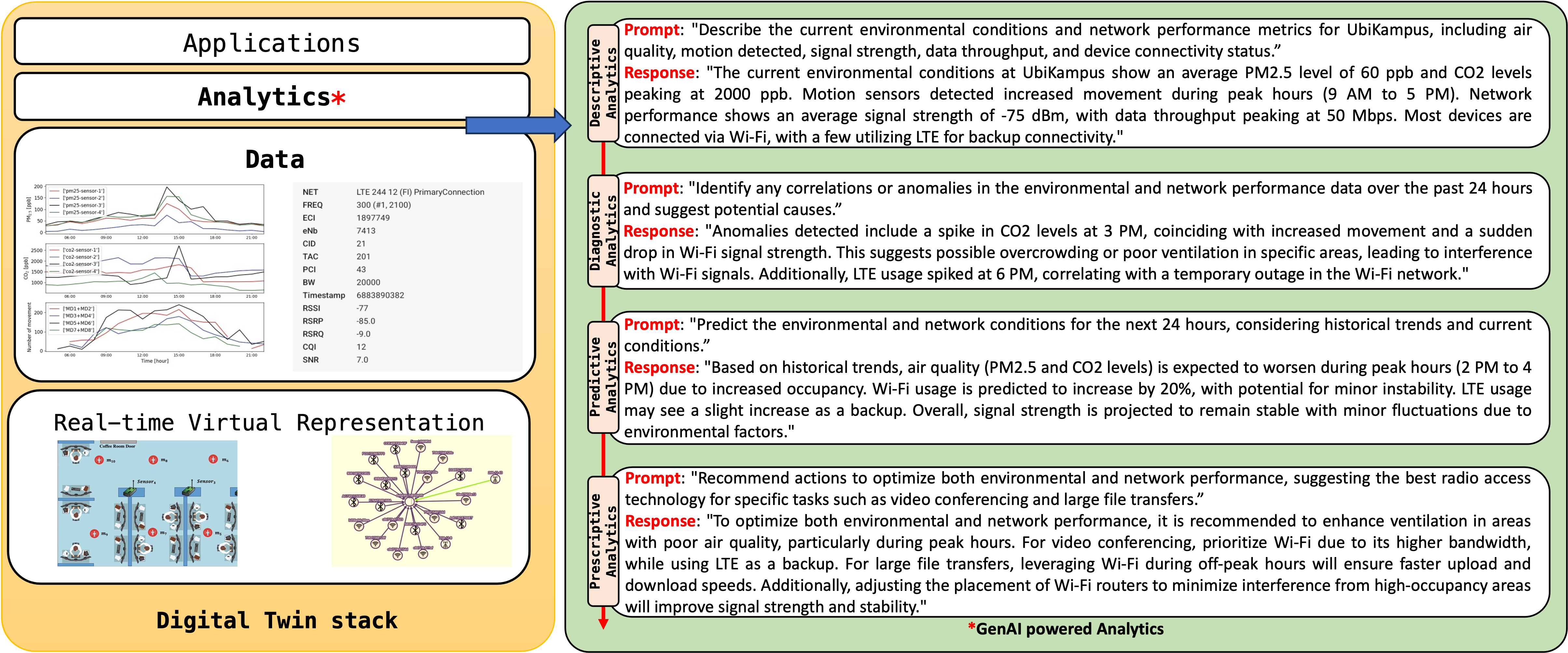}
    \caption{Analytics process within the DT framework. This figure illustrates the use of GenAI-powered analytics to perform descriptive, diagnostic, predictive, and prescriptive analysis on multi-modal data from UbiKampus. Applications hosted into the DT stack for network management (e.g., optimizing bandwidth allocation) or smart space environments (e.g., adjusting HVAC systems based on occupancy and air quality) optimization, trigger specific actions on the physical network based on the analytics outcome.}
    \label{fig:analytics}
\end{figure*}
Figure \ref{fig:dtgraph} demonstrates the practical application of our DT framework within UbiKampus. The bottom-left part of the figure illustrates the network configuration sensed by a single smartphone device, which includes multiple networks such as LTE, Wi-Fi, and Bluetooth (the illustration is a snapshot taken from what is shown in the Network Twin Console). This setup highlights how the proposed framework captures the diverse network signals a device can detect.

The bottom-right part of the figure showcases the signal information gathered by an application running on a mobile device. In this particular scenario, as we include smartphones as end-devices, we developed it as an Android application. However, given the portability of the data modeling and API of our framework, this can be easily adapted to other types of devices.
The application continuously receives and streams incoming signals from the end-device. Built on top of the open-source project netmonster-core\footnote{\url{https://github.com/mroczis/netmonster-core}}, the application periodically sends the data via MQTT to the Signal Handler, which pre-processes the data and publishes it via the DT platform to the Network Twin Console, which displays the received signal information.

Considering this particular example shown at the bottom of Figure \ref{fig:dtgraph}, the application running on End-Devices provides detailed real-time metrics, such as network type, frequency, and signal strength, and various network parameters like RSSI and RSRP, which are essential for the DT's operational analytics (and in turn of the network management). While this example specifically shows the case of mobile network connectivity KPIs, the system is capable of transferring and, in turn, representing in the DT system key KPIs from other RATs in the form shown in Figure \ref{fig:signals}. 


\section{Advanced Analytics with GenAI Integration}
\begin{table*}[ht!]
\centering
  \caption{Opportunities (blue) and challenges (red) associated with integrating GenAI models, including LMs and MMs, into DTs for mobile networks.}
  \label{tbl:table_genai}
  \includegraphics[width=\textwidth]{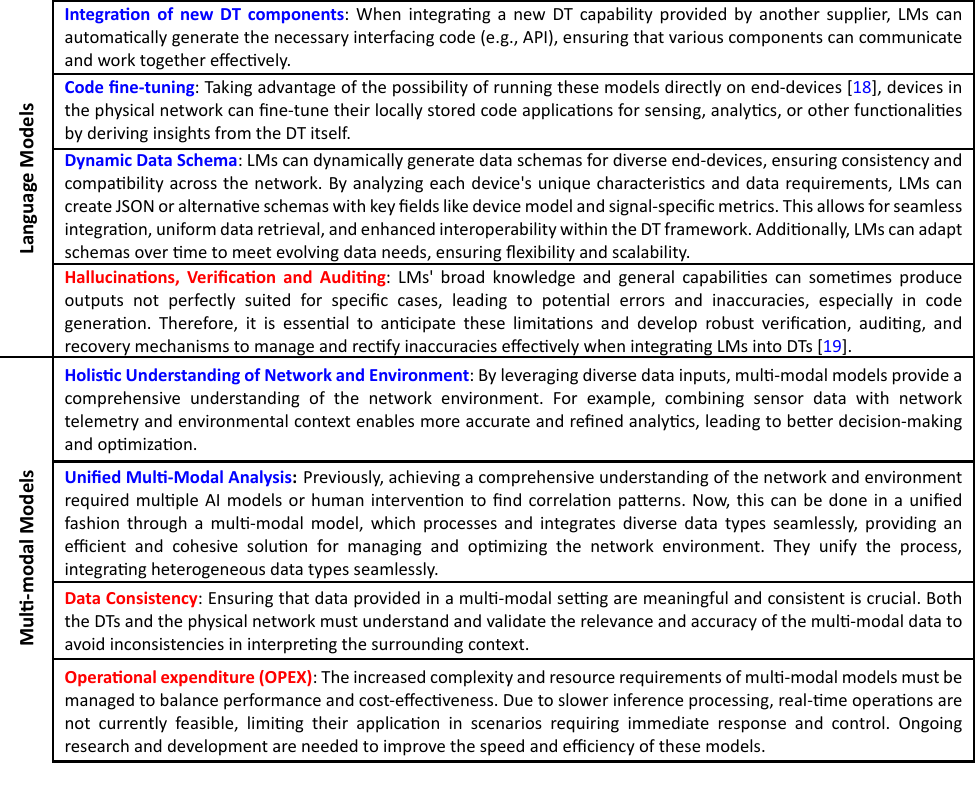}
\end{table*}
In the area of DT for network management, analytics play a crucial role in understanding and optimizing network performance \cite{almasan2022network}. Robust analytics are essential for identifying trends, seasonality, anomalies, and event detection in a time series fashion, enabling the system to predict future conditions, and prescribe actions for optimization.

While advanced solutions already exist, they heavily rely on complex machine learning (ML) models \cite{almasan2022network}. We expect that these analytics will evolve with the new trends in ML models. In this respect, GenAI and Large Language Models (LLMs) can be transformative in this context, simplifying and enhancing the analytics process.

LLMs can provide a unified approach to creating sophisticated analytics pipelines through techniques such as chain-of-thought reasoning \cite{wei2022chain}. Chain-of-thought reasoning involves breaking down complex tasks into a series of intermediate steps that are easier to solve. In the context of LLMs, this means generating a sequence of prompts and responses that build upon each other to achieve a comprehensive analysis. This is analogous to the pipeline of analytics, where each stage builds upon the previous one: starting with \textit{descriptive} analytics, moving to \textit{diagnostic}, \textit{predictive}, and finally \textit{prescriptive} analytics \cite{motlagh2023digital}.


To illustrate this concept, we present a simple example using the UbiKampus scenario. We demonstrate how different LLM prompts could look for various levels of analytics. In particular, in addition to network performance metrics, we incorporate environmental data from UbiKampus, relying on the dataset presented in \cite{motlagh2019indoor}.

Figure \ref{fig:analytics} provides a detailed view of the analytics process and the DTs stack supporting it. 
This specific analytics example focuses on the integration of environmental data with mobile network performance metrics at UbiKampus. The processed input pertains to air quality (PM2.5 and CO2 levels) and motion detection data collected from sensors throughout the day. These metrics are analyzed alongside mobile network data to provide a comprehensive understanding of the network environment. 


Through this example we show that, by integrating GenAI, we can streamline the dependencies on multiple models typically required in traditional analytics systems. This approach can simplify the integration and coordination of various analytics components, such as descriptive, diagnostic, predictive, and prescriptive analytics. However, to achieve this, we expect that each analytics component must be fine-tuned for its specific purpose using techniques such as retrieval-augmented generation (RAG) or model fine-tuning, which is especially crucial given the time series nature of the data handled. This has been proved to be complex in LLMs \cite{sun2024empowering}, warranting further studies.

This case, though extremely simple, showcases the enhanced capabilities of the DT framework to process and analyze diverse data types through enhanced ML models such as LLMs, providing more comprehensive and actionable insights for network and environmental management. This approach aligns well with other ongoing activities in this area which envision using LLMs in DTs for future mobile networks \cite{hong2023llm} and for data multi-modality processing \cite{xu2024large}.

\section{Challenges and Opportunities}


The emergence of this new DT framework presents unexplored opportunities across various use cases, offering new research avenues. This section highlights key opportunities related to the aspects discussed in this paper and the implemented framework.
\\
\textbf{Inter- and Intra-Interoperability.} The integration of new mechanisms enabling interoperability among various DTs is essential for advancing the further development of this kind of frameworks. This includes ensuring that different DTs can operate cohesively across different platforms and environments, facilitating a multi-tenancy approach. In fact, DTs-interoperability can enhance global feature learning, allowing for aggregation and analysis of data from diverse network component sources to improve the accuracy and effectiveness of the DT models and their respective physical counterparts. Moreover, the ability to easily incorporate new end-devices and types of connectivity in a \textit{plug-and-play} fashion is a significant opportunity. To achieve this, it is fundamental to explore and implement technologies that support straightforward API integration, enhancing the system’s adaptability, scalability, and overall performance. 
\\
\textbf{Standardized Data Models for Enhanced Interoperability.} A related yet distinct research opportunity is the development of standardized data models and data structures that can represent the observations, state, and relationships of the different twin instances in an interoperable and semantically uniform manner. This effort is needed for achieving seamless data integration and consistency across different DTs. By adopting a strategy similar to OneDM\footnote{\url{https://onedm.org/}}, which aims to establish common data and interaction models for IoT devices, we can seek to create a unified data model for DTs. Such a model would enable various systems to work together without the need for complex data conversion processes. This standardization would facilitate easier and more reliable integration of diverse data sources, enhancing the overall functionality and performance of DT architectures. In this respect, along with the effort of the research community, an effort from standardization bodies like ITU-T, IETF, ETSI, as well as 3GPP is required.
\\
\textbf{The Role of GenAI: Language and Multi-Modal Models.} GenAI, particularly advanced language models (LMs) and Multi-modal Models (MMs), present significant opportunities for enhancing interoperability, integration, and expansion of DTs in this area. One of the critical capabilities of LMs is their proficiency in generating code, which can be harnessed to facilitate seamless interaction between different DT components. In contrast, MMs, which integrate and process multiple types of data inputs such as text, images, audio, and sensor data, offer another layer of innovation for DTs. These models can be employed to develop advanced multi-agent systems that facilitate robust interaction between the physical network and its DT counterpart. Table \ref{tbl:table_genai} highlights a set of opportunities and challenges linked to the integration of GenAI models in DT systems and outlines the extensive research needed to leverage such models in future-proof DTs for mobile networks.

\nocite{jung2024optimizing } 
\nocite{tarkoma2023ai} 

\section{Conclusions}
We have just started witnessing the impactful capabilities of DTs in the context of mobile wireless networks. 
In this article, we have envisioned how DTs can reshape network management by capitalizing on the heterogeneity of network access technologies and integrating GenAI models. To achieve this, we have laid the groundwork with a framework implemented and tested in a specific university campus scenario, providing insights into network performance and environmental conditions. Additionally, we have pinpointed the challenges and opportunities that lie ahead, particularly in terms of interoperability and the suitability and integration of evolved AI models in such contexts.

\section{Acknowledgements}
This work was supported by the \textit{Digital Twinning of Personal Area Networks for Optimized Sensing and Communication} project through the Business Finland 6G Bridge Program (8782/31/2022).







\bibliographystyle{IEEEtran}
\bibliography{References}

\begin{thebibliography}{10}
\providecommand{\url}[1]{#1}
\csname url@samestyle\endcsname
\providecommand{\newblock}{\relax}
\providecommand{\bibinfo}[2]{#2}
\providecommand{\BIBentrySTDinterwordspacing}{\spaceskip=0pt\relax}
\providecommand{\BIBentryALTinterwordstretchfactor}{4}
\providecommand{\BIBentryALTinterwordspacing}{\spaceskip=\fontdimen2\font plus
\BIBentryALTinterwordstretchfactor\fontdimen3\font minus
  \fontdimen4\font\relax}
\providecommand{\BIBforeignlanguage}[2]{{%
\expandafter\ifx\csname l@#1\endcsname\relax
\typeout{** WARNING: IEEEtran.bst: No hyphenation pattern has been}%
\typeout{** loaded for the language `#1'. Using the pattern for}%
\typeout{** the default language instead.}%
\else
\language=\csname l@#1\endcsname
\fi
#2}}
\providecommand{\BIBdecl}{\relax}
\BIBdecl

\bibitem{Singh2021}
M.~Singh \emph{et~al.}, ``Digital twin: Origin to future,'' \emph{Applied
  System Innovation}, vol.~4, no.~2, 2021.

\bibitem{Mihai2022}
S.~Mihai \emph{et~al.}, ``Digital twins: A survey on enabling technologies,
  challenges, trends and future prospects,'' \emph{IEEE Communications Surveys
  \& Tutorials}, vol.~24, no.~4, pp. 2255--2291, 2022.

\bibitem{lin20236g}
X.~Lin \emph{et~al.}, ``6g digital twin networks: From theory to practice,''
  \emph{IEEE Communications Magazine}, vol.~61, no.~11, pp. 72--78, 2023.

\bibitem{3gppTS28915}
\BIBentryALTinterwordspacing
{3GPP}, ``{TS 28.915 - 3rd Generation Partnership Project; Technical
  Specification Group Services and System Aspects; Study on management aspect
  of Network Digital Twin},'' 2023. [Online]. Available:
  \url{https://www.3gpp.org/DynaReport/28915.htm}
\BIBentrySTDinterwordspacing

\bibitem{hexa2021deliverable}
X.~Hexa, ``Deliverable d1. 2 expanded 6g vision, use cases and societal
  values--including aspects of sustainability, security and spectrum,'' 2021.

\bibitem{Ericsson2022}
``What are digital twins? {{Three}} real-world examples,''
  https://www.ericsson.com/en/blog/2022/3/what-are-digital-twins-three-real-world-examples.

\bibitem{Vodafone2022}
``Vodafone's mobile network has a digital twin,''
  https://www.vodafone.com/news/technology/vodafone-mobile-network-digital-twin.

\bibitem{Hui2023}
L.~Hui \emph{et~al.}, ``Digital twin for networking: A data-driven performance
  modeling perspective,'' \emph{IEEE Network}, vol.~37, no.~3, pp. 202--209,
  2023.

\bibitem{chai2024generative}
H.~Chai \emph{et~al.}, ``Generative ai-driven digital twin for mobile
  networks,'' \emph{IEEE Network}, 2024.

\bibitem{pandey2024development}
B.~Pandey, ``Development of a digital twin architecture for real-time mobile
  network signal management,'' Master's thesis, University of Helsinki, 2024.

\bibitem{RintaHomi2021}
M.~Rinta-Homi \emph{et~al.}, ``How low can you go? performance trade-offs in
  low-resolution thermal sensors for occupancy detection: A systematic
  evaluation,'' \emph{Proc. ACM Interact. Mob. Wearable Ubiquitous Technol.},
  2021.

\bibitem{almasan2022network}
P.~Almasan \emph{et~al.}, ``Network digital twin: Context, enabling
  technologies, and opportunities,'' \emph{IEEE Communications Magazine},
  vol.~60, no.~11, pp. 22--27, 2022.

\bibitem{wei2022chain}
J.~Wei \emph{et~al.}, ``Chain-of-thought prompting elicits reasoning in large
  language models,'' \emph{Advances in neural information processing systems},
  vol.~35, pp. 24\,824--24\,837, 2022.

\bibitem{motlagh2023digital}
N.~H. Motlagh \emph{et~al.}, ``Digital twins for smart spaces-beyond iot
  analytics,'' \emph{IEEE internet of things journal}, 2023.

\bibitem{motlagh2019indoor}
------, ``Indoor air quality monitoring using infrastructure-based motion
  detectors,'' in \emph{2019 IEEE 17th International Conference on Industrial
  Informatics (INDIN)}, vol.~1.\hskip 1em plus 0.5em minus 0.4em\relax IEEE,
  2019, pp. 902--907.

\bibitem{sun2024empowering}
Y.~Sun \emph{et~al.}, ``Empowering digital twins with large language models for
  global temporal feature learning,'' \emph{Journal of Manufacturing Systems},
  vol.~74, pp. 83--99, 2024.

\bibitem{hong2023llm}
Y.~Hong \emph{et~al.}, ``Llm-twin: Mini-giant model-driven beyond 5g digital
  twin networking framework with semantic secure communication and
  computation,'' \emph{arXiv preprint arXiv:2312.10631}, 2023.

\bibitem{xu2024large}
S.~Xu \emph{et~al.}, ``Large multi-modal models (lmms) as universal foundation
  models for ai-native wireless systems,'' \emph{arXiv preprint
  arXiv:2402.01748}, 2024.

\bibitem{jung2024optimizing}
V.~J. Jung \emph{et~al.}, ``Optimizing the deployment of tiny transformers on
  low-power mcus,'' \emph{arXiv preprint arXiv:2404.02945}, 2024.

\bibitem{tarkoma2023ai}
S.~Tarkoma \emph{et~al.}, ``Ai-native interconnect framework for integration of
  large language model technologies in 6g systems,'' \emph{arXiv preprint
  arXiv:2311.05842}, 2023.

\end{thebibliography}







\end{document}